\documentclass[conference]{IEEEtran}
\IEEEoverridecommandlockouts
\usepackage{graphicx}   
\usepackage{textcomp}
\usepackage{caption}
  \usepackage{comment}
  \usepackage{lipsum}
\usepackage{xcolor}
\def\BibTeX{{\rm B\kern-.05em{\sc i\kern-.025em b}\kern-.08em
    T\kern-.1667em\lower.7ex\hbox{E}\kern-.125emX}}
\usepackage{amsmath,amssymb,amsfonts}
\usepackage{amsmath}  
\usepackage{amssymb} 
\usepackage{multirow}
\usepackage{cite}   
\usepackage{breqn}  
\renewcommand{\baselinestretch}{1}
\usepackage{algcompatible}
\usepackage{algorithm} 

\usepackage{caption}
\newcommand{\qed}{\nobreak \ifvmode \relax \else
      \ifdim\lastskip<1.5em \hskip-\lastskip
      \hskip1.5em plus0em minus0.5em \fi \nobreak
      \vrule height0.75em width0.5em depth0.25em\fi}
      \usepackage[caption=false,font=footnotesize]{subfig}

\graphicspath{{./figures/}} 
\begin{document}
%
%
%
%
\title{Future Ultra-Dense LEO Satellite Networks: A Cell-Free Massive MIMO Approach}
\author{\IEEEauthorblockN{Mohammed~Y.~Abdelsadek\IEEEauthorrefmark{1},~~Halim~Yanikomeroglu\IEEEauthorrefmark{1}~~and Gunes~Karabulut~Kurt\IEEEauthorrefmark{4}}
\IEEEauthorblockA{\IEEEauthorrefmark{1}Department of Systems and Computer Engineering,  Carleton University, Ottawa, ON K1S 5B6, Canada}
\IEEEauthorblockA{\IEEEauthorrefmark{4} Department of Electronics and Communications Engineering, Istanbul Technical University, Istanbul, 34469, Turkey}
\IEEEauthorblockA{Emails: mohammed.abdelsadek@carleton.ca,~halim@sce.carleton.ca,~ gkurt@itu.edu.tr}}

%
\maketitle
\begin{abstract}
Low Earth orbit (LEO) satellite networks (SatNets) are envisioned to play a crucial role in providing global and ubiquitous connectivity efficiently. Accordingly, in the coming years, thousands of LEO satellites will be launched to create ultra-dense LEO mega-constellations, and the Third Generation Partnership Project (3GPP) is working on evolving fifth-generation (5G) systems to support such non-terrestrial networks (NTN). However, many challenges are associated with the deployment of LEOs from communications and networking perspectives. In this paper, we propose a novel cell-free massive multiple-input multiple-output (CF-mMIMO) based architecture for future ultra-dense LEO SatNets. We discuss various aspects of network design, such as duplexing mode, pilot assignment, beamforming, and handover management. In addition, we propose a joint optimization framework for the power allocation and handover management processes to maximize the network throughput and minimize the handover rate while ensuring quality-of-service (QoS) satisfaction for users. To the best of our knowledge, this is the first work to introduce and study CF-mMIMO-based LEO SatNets. Extensive simulation results demonstrate the superior performance of the proposed architecture and solutions compared to those of conventional single-satellite connectivity and handover techniques from the literature.
\end{abstract}
%
%
%

%
\section{Introduction}
\label{sec:introduction}
%
%
Despite the remarkable advancements in wireless connectivity and the advent of fifth-generation (5G) networks, approximately half of the world still has limited or no Internet access \cite{3GPPTR22-822}. To address this lack, satellite networks (SatNets) can play a significant role in providing global and ubiquitous connectivity, which has yielded a resurgence of interest in exploiting satellites to this end. Compared to geostationary Earth orbit (GEO) satellites, low Earth orbit (LEO) satellites can be deployed at altitudes as low as $300$ km. This overcomes the long propagation delay and high path loss challenges associated with GEO satellites and provides a more favourable communications and networking environment. Accordingly, players such as SpaceX, OneWeb, and Telesat are planning to launch thousands of LEO satellites in the coming years to create ultra-dense constellation deployments. Moreover, the Third Generation Partnership Project
(3GPP) has included several study items in Releases 15 and 16 to examine the support of non-terrestrial networks (NTN) in the 5G new radio (NR) \cite{3GPPTR38-811,3GPPTR38-821}. A work item has been approved for the standardization of 5G NR-NTN in Release 17 \cite{3GPPRP-193234}.


%
Although LEO SatNets provide several advantages over GEOs, the high mobility of LEO satellites relative to ground user terminals (UTs) poses many challenges for the allocation of radio resources and handover management processes. For example, every handover process is associated with signaling overhead, processing delay, throughput losses, and data forwarding. To address these challenges, several techniques have been considered in the literature that aim to ensure users' quality-of-service (QoS) satisfaction while maximizing network utility. One of the most promising techniques is massive multiple-input multiple-output (MIMO) technology in LEO SatNets. In \cite{you2020massive}, the authors studied the deployment of massive MIMO in LEO SatNets, assuming the LEO satellites were equipped with uniform planar arrays of antennas to serve ground UTs. In so doing, the authors investigated precoding and user grouping utilizing statistical channel state information. However, they did not consider the distributed deployment of antenna systems. In \cite{erdogan2020site}, the authors studied the physical layer performance of the optical SatNets utilizing site diversity to connect to multiple ground stations. The authors in \cite{feng2020satellite} modeled ground gateway stations and visible LEO satellites as a bipartite graph, assuming that this information was known by the ground gateways. They then proposed a maximum matching based solution to select the satellites that could be connected to every ground station considering basic MIMO concepts to deal with this multi-connectivity. In \cite{goto2018leo}, the authors studied the capacity of LEO-MIMO systems considering the Doppler shift and allocating different channels to data and control signals. However, these studies considered basic MIMO models to describe the connectivity of ground terminals with multiple satellites without investigating the details of the network architecture, channel estimation, precoding, or interference between users.   

Due to the density of LEO constellations, the visibility of more than a dozen LEO satellites at the same time by UTs will be a reality. This offers an unprecedented opportunity to employ architectures such as cell-free massive MIMO (CF-mMIMO), which can yield substantial performance improvements in future LEO SatNets.  CF-mMIMO \cite{ngo2017cell} is a recent technology introduced for next-generation terrestrial networks that uses techniques from network MIMO and massive MIMO for large spectral efficiency and network flexibility improvements. In this paper, we propose an LEO satellite network architecture based on a CF-mMIMO framework. Specifically, we discuss network topology, inter-satellite links (ISLs), duplexing mode, pilot assignment, beamforming, and handover management. We jointly optimize the power allocation and handover management processes such that the aggregate data rate of UTs is maximized and their handover rate is minimized while considering their different QoS requirements, based on the proposed CF-mMIMO architecture. For this purpose, we investigate the channel model, uplink training and channel estimation, downlink data transmission, and problem formulation. Finally, extensive simulations are conducted to evaluate the performance of the proposed architecture and solutions, and we compare the performance with that of conventional approaches and architectures. To the best of our knowledge, this is the first work to introduce and study CF-mMIMO in LEO SatNets.

The paper is organized as follows. In Section \ref{sec:architecture}, we introduce the CF-mMIMO-based LEO satellite network architecture and investigate various aspects of the network, i.e., time-division duplexing (TDD) operation, pilot assignment, beamforming, and handover management. Then, we discuss the joint power allocation and handover management problem in Section \ref{sec:JointPAandHM}. In Section \ref{sec:Results}, we present and discuss the results of the simulations to evaluate the performance of the proposed CF-mMIMO architecture by comparing it with that of traditional single satellite connectivity. Finally, conclusions are presented in Section \ref{sec:Conclusions}.
%
\section{CF-mMIMO-Based LEO SatNets}
\label{sec:architecture}
\subsection{Architecture}
\begin{figure}
\centering     
\includegraphics[width=80mm]{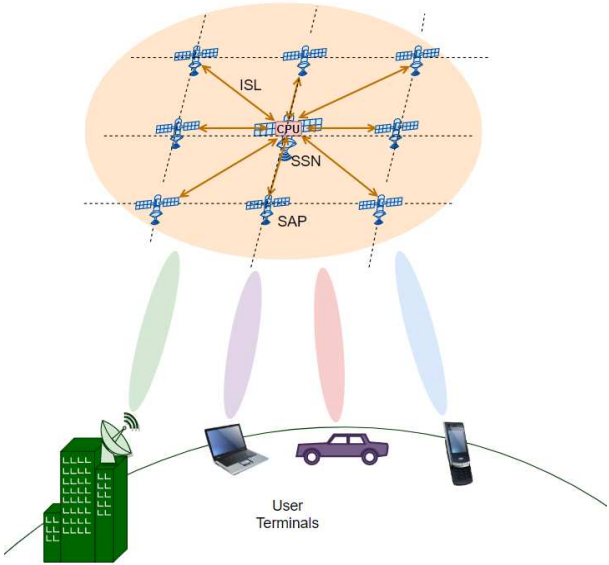}
\caption{Proposed CF-mMIMO-based LEO SatNet architecture.}
\label{fig:architecture}
\end{figure}

Fig. \ref{fig:architecture} shows the proposed CF-mMIMO-based LEO SatNet architecture with the satellites divided into clusters. To be consistent with the CF-mMIMO terminology, each satellite in the cluster is called a satellite access point (SAP). We should add that for satellite nodes with multiple antennas, each antenna is considered a separate SAP. The SAPs are connected to a central processing unit (CPU) through ISLs for backhaul purposes. This CPU can be deployed on a central satellite with more advanced computing capabilities. We call these satellites super satellite nodes (SSNs). The effect of the mobility of the satellites on this topology can be negligible due to the fact that all these satellites move as a group, and the relative speed between them can be neglected. The selection of the number of SSNs and the number of SAPs in each cluster can be optimized on the basis of different performance objectives and the associated deployment costs.

The exchange of the information between the SAPs and the CPU can be restricted to payload data and power control coefficients to reduce the backhaul signaling, as discussed in \cite{ngo2017cell}. However, backhaul signaling is not considered a challenge in this case as in terrestrial networks. This is because ISLs can use high-speed free-space optical (FSO) communications \cite{chaudhry2020free}, which enables information to be exchanged at very high data rates and low latency.  
%

\subsection{TDD Operation}
To exploit the reciprocity of the uplink (UL) and downlink (DL) channels, TDD is the adopted duplexing mode in CF-mMIMO systems \cite{ngo2017cell}. For channel estimation, both UL and DL pilots can be used \cite{interdonato2019ubiquitous}. However, considering only UL pilots is adopted in most CF-mMIMO studies, e.g., \cite{ngo2017cell,chen2018channel,bjornson2019making}. This is suitable for SatNets because the users may not need to estimate their effective channel gain, and to use most of the TDD frame for data transmission. 

The channel coherence interval is defined as the time-frequency interval during which the channel characteristics can be considered static. This coherence interval depends on the channel condition, the mobility of the satellite and UT, and the carrier frequency. The channel uses are determined in accordance with the coherence time, $\tau_c$, and divided into three segments: the initial $\tau^p_u$ samples are used for UL pilot transmission; the next $\tau^d_u$ samples are used for UL data transmission; and the last $\tau^d_d$ samples are reserved for DL data transmission. The guard intervals are excluded from this coherence time interval. Utilizing the UL pilots, all the UL channels are estimated at the SAPs locally without forwarding them to the CPU. This supports the scalability of the network since the signaling overhead is independent of the number of SAPs. Due to reciprocity, these channel estimates are valid for the DL direction as well. Therefore, the estimated channels are used for DL data precoding and UL data detection.
%

\subsection{Pilot Assignment}
UTs can be assigned mutually orthogonal UL pilots to minimize the interference between them. However, this requires the number of UL training samples, $\tau^p_u$, to be more than the connected UTs, which is difficult in SatNets due to the high number of connected UTs. Therefore, every subset of the UTs can be assigned one pilot from the mutually orthogonal pilot set. This results in what is known as pilot contamination, which needs to be taken into consideration when designing the resource allocation procedure. The pilot assignment can be implemented locally at the SAPs in a distributed manner or centrally at the CPU. The pilot assignment information can be transmitted to the UTs in the random access channel during the random access process.

\subsection{Beamforming}
Maximum ratio processing (i.e., conjugate beamforming in the DL direction and matched filtering in the UL), can be employed to exploit the distributed channel estimation at the SAPs. This is considered one of the major benefits of using CF-mMIMO, as it has the potential to reduce the computational complexity and the required backhaul signaling between the SAPs and the CPU \cite{ngo2017cell}. However, other beamforming techniques can be used, such as zero-forcing (ZF) and minimum mean square error (MMSE) \cite{nayebi2017precoding,bjornson2019making}. These centralized techniques can improve performance, but they require more backhaul signaling between the SAPs and the CPU.


\subsection{Handover Management}
In the case of single-satellite association, when the signal level is below a certain threshold, the link is switched to the next LEO satellite in the cone visibility of the UT. This can be accomplished by using the satellite reference signals that are broadcast by the satellite. In this case, the service time, which is the connection time without
handover interruption, is limited by the satellite visibility. However, using the proposed CF-mMIMO architecture, the UT is connected to a cluster of satellites or SAPs. Consequently, the service time is limited by the visibility of the whole cluster, which is longer than that of a single satellite.

In addition, power allocation can be adjusted such that the service times of the ground UTs are maximized. This is because the UTs are served by all SAPs in the serving cluster. Therefore, the cooperative transmission of those SAPs can compensate for the signal level decaying due to the movement of the satellites. Moreover, in the UL direction, the data is decoded on the basis of the received signals by all SAPs in the cluster. Nevertheless, a cluster handover is required to switch to the next cluster when resource allocation cannot satisfy the UT's minimum required data rate level. This can be detected while allocating the radio resources (i.e., the transmit power), as will be discussed in Section \ref{sec:JointPAandHM}. Besides, due to the fact that the next cluster is known, the handover decision can be confirmed by the next cluster that detects the UL pilot from the  ground UT by its edge SAPs. This confirmation can be used to avoid false handover decisions.
%
\section{Joint Power Allocation and Handover Management}
\label{sec:JointPAandHM}
Power control plays an essential role in optimizing the cooperative transmission and reception of the SAPs to maximize the network throughput and ensure user satisfaction. The power allocation process should consider the interference between the UTs, the pilot assignment, and the achievable data rates. In addition, power allocation can be optimized to maximize the service time of the UTs to minimize their handover rate. In this section, we investigate the problem of joint power allocation and handover management for LEO SatNets based on the proposed CF-mMIMO architecture. For this purpose, we start by discussing the channel model and estimation. Then, we derive the achievable data rates and formulate the optimization problem for joint power allocation and handover management.



\subsection{Channel Model}
We consider a cluster of LEO satellites that includes a set of $M$ SAPs indexed by $\mathcal{M}=\{1,~2,\cdots,~m,\cdots,~M\}$. This cluster serves a set of single-antenna ground UTs set, indexed by $\mathcal{K}=\{1,~2,\cdots,~k,\cdots,~K\}$. We assume that the channel conditions are static in a coherence time interval of $\tau_c$ samples. Due to the strong LoS component between the UTs and SAPs, the channel between the $k$th UT and the $m$th SAP is modeled as Rician and can be calculated by \cite{ngo2018performance}
\begin{align}
\label{eq:channelCoeff1}
h_{m,k}=\sqrt{L_{m,k}}\left(\sqrt{\frac{\kappa_{m,k}}{\kappa_{m,k}+1}} h'_{m,k}+\sqrt{\frac{1}{\kappa_{m,k}+1}}h''_{m,k} \right),
\end{align}
where $\kappa_{m,k}$ is the Rician K-factor, $h'_{m,k}$ and $h''_{m,k}$ represent the LoS and non-LoS (NLoS) components, respectively. The large scale fading and losses are represented by $L_{m,k}=10^{-(L^{dist}_{m,k}+L^{shad}_{m,k}+L^{angl}_{m,k})/10}$, where $L^{dist}_{m,k}$ is the power loss (in dB) due to distance between UT $k$ and SAP $m$, $L^{shad}_{m,k} \sim\mathcal{N}(0,\sigma_{sh}^2)$ is the shadowing attenuation (in dB) with variance $\sigma_{sh}^2$, and $L^{angl}_{m,k}$ is the loss due to the boresight angle and can be calculated (in dB) by \cite{li2020hierarchical}
\begin{align}
L^{angl}_{m,k}=-10 \log_{10} \left( \cos(\theta_{m,k})^{\eta} \frac{32 \log 2}{2\left( 2~\text{arccos}(\sqrt[\eta]{0.5}) \right)^2} \right),
\end{align}
where $\theta_{m,k}$ is the boresight angle between the $k$th UT and the $m$th SAP, and $\eta$ is the antenna roll-off factor determining the coverage radius, assuming that the aperture efficiency is unity.

We assume that the NLoS component is a Rayleigh random variable, i.e., $h''_{m,k}\sim \mathcal{C}\mathcal{N}(0,1)$. The LoS component is given by $h'_{m,k}=e^{j\phi_{m,k}}$, where $\phi_{m,k}\sim \mathcal{U}[-\pi,\pi]$ is a uniform random variable that represents the phase shift due to the mobility of the SAPs and UTs and propagation delay. Therefore, this is an important factor and should be taken into account while considering CF-mMIMO in LEO satellites that are characterized by their high mobility relative to ground terminals. 

For simplicity, we rewrite (\ref{eq:channelCoeff1}) as follows:
\begin{align}
\label{eq:channelCoeff2}
h_{m,k}=\sqrt{\beta_{m,k}}e^{j\phi_{m,k}} + \tilde{h}_{m,k},
\end{align}
where $\beta_{m,k} = \kappa_{m,k}L_{m,k}/(\kappa_{m,k}+1)$, $\tilde{h}_{m,k}\sim \mathcal{C}\mathcal{N}(0, \lambda_{m,k})$, $\lambda_{m,k}=L_{m,k}/(\kappa_{m,k}+1)$, and $\beta_{m,k}\in \mathbb{R}$. Therefore, $\beta_{m,k}$ and $\lambda_{m,k}$ are changing slowly and can be known a priori.
%
\subsection{Uplink Training and Channel Estimation}
As discussed in Section \ref{sec:architecture}, TDD is adopted as the duplexing technique. Therefore, to estimate the UL channels at the SAPs, every UT transmits a pilot signal on the initial $\tau_u^p$ samples of the coherence block. Since we assume that the number of UTs is greater than the number of mutual orthogonal pilots, i.e., $K>\tau_u^p$, every subset of UTs is assigned the same pilot. The subset of UTs that is assigned the same pilot as UT $k$ is denoted by $\mathcal{C}_k$. We define $\sqrt{q_k}\psi_k\in \mathbb{C}^{\tau_u^p\times 1}$ as the $\tau_u^p$-length pilot sequence transmitted by the $k$th UT, where $q_k$ is the pilot power and $\psi_k^H\psi_k=||\psi_k||^2=\tau_u^p$. Therefore, the received signal vector at the $m$th SAP, $\textbf{y}_m^p\in \mathbb{C}^{\tau_u^p\times 1}$, from all $K$ UTs' pilot transmissions is given by
\begin{align}
\textbf{y}_m^p=\sum_{k=1}^K \sqrt{q_k}h_{m,k}\psi_k + \textbf{n}_m^p,
\end{align}
where $\textbf{n}_m^p\sim \mathcal{CN}(\textbf{0}_{\tau_u^p},\sigma_{n^p}^2\textbf{I}_{\tau_u^p})$ is the additive white Gaussian noise (AWGN) vector.

To estimate the UL channel of UT $k$, sufficient statistics are derived from the received signal by calculating the inner product between the received signal vector, $\textbf{y}_m^p$, and $\psi_k$, as follows:
\begin{align}
y_{m,k}^p&=\psi_k^H \textbf{y}_m^p=\sum_{k'=1}^K \sqrt{q_{k'}}h_{m,k'}\psi_k^H \psi_{k'}+\psi_k^H \textbf{n}_m^p\\
&= \sqrt{q_k}\tau_u^ph_{m,k}+\sum_{k'\in\mathcal{C}_k\backslash \{k\}} \sqrt{q_{k'}}h_{m,k'}\tau_u^p+\psi_k^H \textbf{n}_m^p.
\end{align}
This is because
\begin{align}
\psi_k^H \psi_{k'}= \begin{cases}
\tau_u^p, & k'\in \mathcal{C}_k\\
0, & \text{otherwise}\\
\end{cases}.
\end{align}

This statistic can be used to estimate the uplink channel, $h_{m,k}$, at the $m$th SAP using techniques such as MMSE and linear MMSE (LMMSE) estimators. We assume that a phase-aware MMSE channel estimator is used. Therefore, the estimated UL channel can be given by \cite{ozdogan2019performance} 
\begin{align}
\label{eq:estimatedChannel}
\hat{h}_{m,k}&=\sqrt{\beta_{m,k}}e^{j\phi_{m,k}}+\frac{\sqrt{q_k}\lambda_{m,k}(y^p_{m,k}-\bar{y}^p_{m,k})}{\gamma_{m,k}},\\
\bar{y}^p_{m,k}&=\sum_{k'\in\mathcal{C}_k} \sqrt{q_{k'}} \tau_u^p \sqrt{\beta_{m,k'}}e^{j\phi_{m,k'}},\\
\gamma_{m,k}&= \sum_{k'\in \mathcal{C}_k}q_{k'}\tau_u^p \lambda_{m,k'}+\sigma_{n^p}^2,
\end{align}
with the following statistics
\begin{align}
\mathbb{E}\{\hat{h}_{m,k}|\phi_{m,k}\}&=\sqrt{\beta_{m,k}}e^{j\phi_{m,k}}\\
\text{Var}\{\hat{h}_{m,k}|\phi_{m,k}\}&= \frac{q_k\tau_u^{p}\lambda_{m,k}^2}{\gamma_{m,k}}.
\end{align}
%
\subsection{Downlink Data Transmission}
Since most of the traffic is in the DL direction, we focus on the DL power allocation problem. In the DL, the SAPs transmit the same symbol to the UT in a cooperative manner. We assume that the symbol to be sent to UT $k$ is $s_k\in \mathbb{C}$. Every symbol is precoded by a precoding vector $\textbf{v}_k=[v_{1,k},v_{2,k},\cdots,v_{M,k}]^T$, where $v_{m,k}\in \mathbb{C}$. Therefore, if the signal vector to be sent to the $K$ UTs is $\textbf{s}=[s_1,s_2,\cdots,s_K]^T$, then the signal vector to be transmitted by the $M$ SAPs is given by
\begin{align}
\textbf{x}&=\textbf{V}\textbf{s}= \textbf{v}_1s_1+\textbf{v}_2s_2+\cdots+\textbf{v}_Ks_K,
\end{align} 
where $\textbf{V}=[\textbf{v}_1,\textbf{v}_2,\cdots,\textbf{v}_K]$ is an $M\times K$ matrix. 

It follows from this that the signal received by the $k$th UT can be calculated as follows:
\begin{align}
y_k&=\textbf{h}_k^H\textbf{x}\\
&=\textbf{h}_k^H\textbf{v}_ks_k+\sum_{k'\in \mathcal{K}\backslash k}\textbf{h}_k^H\textbf{v}_{k'}s_{k'}+n_k,
\end{align}
where $\textbf{h}_k=[h_{1,k},h_{2,k},\cdots,h_{M,k}]^T$ and $n_k\sim\mathcal{CN}(0,\sigma^2_n)$ is the AWGN noise. Therefore, the signal-to-interference and noise ratio (SINR) can be calculated by
\begin{align}
\label{eq:SINR}
\text{SINR}_k=\frac{|\mathbb{E}\{\textbf{v}_k^H\textbf{h}_k\}|^2}{\sum_{k'\in \mathcal{K}\backslash k}\mathbb{E}\{|\textbf{v}_{k'}^H\textbf{h}_k|^2\}+\sigma_n^2},
\end{align}
where $\mathbb{E}\{\cdot\}$ is the expectation.

In this study, we adopt coherent beamforming as the technique used to minimize the interference between the UTs. Therefore, the precoding coefficient for the $k$th UT and $m$th SAP is $v_{m,k}=\sqrt{\frac{p_{m,k}}{\mathbb{E}\{|\hat{h}_{m,k}|^2\}}}\hat{h}_{m,k}$, where $p_{m,k}$ is a power scaling factor and $\hat{h}_{m,k}$ is the estimated UL channel, which is valid for the DL direction by virtue of channel reciprocity. That is, the precoding vector for the $k$th UT is given by
\begin{align}
\label{eq:precodingVector}
\textbf{v}_k=\textbf{P}_k^{1/2}\hat{\textbf{h}}_k,
\end{align}
where $\textbf{P}_k=\text{diag}\left(\frac{p_{1,k}}{\mathbb{E}\{|\hat{h}_{1,k}|^2\}},\frac{p_{2,k}}{\mathbb{E}\{|\hat{h}_{2,k}|^2\}},\cdots,\frac{p_{M,k}}{\mathbb{E}\{|\hat{h}_{M,k}|^2\}}\right)$ and $\hat{\textbf{h}}_k=[\hat{h}_{1,k},\hat{h}_{2,k},\cdots,\hat{h}_{M,k}]^T$ .

Accordingly, by using coherent beamforming as in (\ref{eq:precodingVector}),  phase-aware MMSE channel estimation as in (\ref{eq:estimatedChannel}), and the SINR in (\ref{eq:SINR}), the achievable DL data rate (in bps/Hz) of the $k$th UT served by this cluster of SAPs can be calculated as in (\ref{eq:DLDataRate}), at the top of next page, based on the discussion in \cite{ozdogan2019performance}, where
\begin{figure*}[t]
\begin{align}
\label{eq:DLDataRate}
R_k=\frac{\tau_d^d}{\tau_c}\log_2\left(  1+ \frac{|\text{tr}(\textbf{P}_k^{1/2}\textbf{W}_k)|^2}{\sum_{k'=1}^K\text{tr}(\textbf{P}_{k'}\textbf{A}'_k\textbf{W}_{k'}) + \sum_{k'\in\mathcal{C}_k\backslash k} q_kq_{k'}(\tau_u^p)^2|\text{tr}(\textbf{P}_{k'}^{1/2}\textbf{A}_k\textbf{G}_{k'}\textbf{A}_{k'})|^2-\text{tr}(\textbf{P}_k\textbf{B}_k^2)+\sigma_n^2} \right)
\end{align}
\end{figure*}
\begin{align}
\textbf{A}_k&=\text{diag}(\lambda_{1,k},\lambda_{2,k},\cdots,\lambda_{M,k}),\\
\textbf{A}'_k&=\text{diag}(\lambda'_{1,k},\lambda'_{2,k},\cdots,\lambda'_{M,k}),\\
\lambda'_{m,k} &= \lambda_{m,k}+\beta_{m,k},\\
\textbf{B}_k &= \text{diag}(\beta_{1,k},\beta_{2,k},\cdots,\beta_{M,k}),\\
\textbf{W}_k &= q_k\tau_u^p \textbf{A}_k\textbf{G}_k\textbf{A}_k + \textbf{B}_k,\\
\textbf{G}_k &= \text{diag}(\gamma_{1,k},\gamma_{2,k},\cdots,\gamma_{M,k})^{-1}.
\end{align}
\subsection{Problem Formulation}
To optimize the power allocation of SAPs such that the cluster throughput is maximized and the handover rate is minimized, we formulate the power allocation problem as a multi-objective optimization problem. We do this such that the objective functions to be maximized are the UTs' aggregate data rate and their service time before being switched to another cluster. For the latter, we maximize the number of UTs served with guaranteed minimum data rate based on their link conditions. When the link condition does not allow the optimized power allocation to serve the UT with the minimum required data rate, a handover request is issued. A handover decision can be taken when this repeats and the visibility of the UT is confirmed by the next serving cluster. 

To deal with this multi-objective optimization problem, we construct a weighted sum of the two objectives to combine the two conflicting objectives in a single function. That is, the power allocation problem at the $t$th time slot can be formulated as follows:
\begin{align}
	\label{eq:OP1}
	&\max_{p_{m,k},I_k}~~ (1-\alpha) \sum_{k=1}^K R_k(t)I_k(t)+\alpha \sum_{k=1}^K I_k(t)\\	
	\tag{\theequation a}\label{eq:OP1a}\mathrm{s.t.}~~&R_k(t) \geq R_k^{\text{min}}I_k(t),~\forall k\in \mathcal{K}\\	
	\tag{\theequation b}\label{eq:OP1b} & \sum_{k=1}^K p_{m,k}\leq P_m^{\text{max}}, ~\forall m\in \mathcal{M}\\
	\tag{\theequation c}\label{eq:OP1c} &I_k(t) \in \{0,1\}, ~\forall k\in \mathcal{K}\\
	\tag{\theequation d}\label{eq:OP1d} & p_{m,k}\geq 0, ~\forall m\in \mathcal{M},~k\in\mathcal{K},
\end{align}
where $\alpha$ is a weighting coefficient that combines the two competing objectives and can be used to prioritize them. $R_k(t)$ is the data rate of the $k$th UT during the $t$th time slot based on its channel conditions and power allocation during that time slot, as given in (\ref{eq:DLDataRate}). $I_k(t)$ is an indicator variable that indicates whether the $k$th UT can be served by the cluster during the $t$th time slot with an acceptable data rate by optimizing the power allocation. This indicator variable is important here to adapt with the mobility of the cluster by excluding the UTs that cannot be served by the cluster due to distance and should be served by another cluster. Constraint (\ref{eq:OP1a}) is used to ensure that the UTs served satisfy their minimum rate level, where $R_k^{\text{min}}$ is the required minimum rate of UT $k$. Constraint (\ref{eq:OP1b}) is expressed to ensure that the total power scaling factors of every SAP are within the required range, where $P_m^{\text{max}}$ is the maximum total value for the $m$th SAP. The binary value of the indicator variable, $I_k(t)$, and the non-negative value of $p_{u,k}$ are imposed by constraints (\ref{eq:OP1c}) and (\ref{eq:OP1d}), respectively.

This optimization problem can be modeled as a mixed-integer non-linear program (MINLP), which is generally an NP-hard problem due to its combinatorial behaviour. Therefore, an exponential computational complexity is required to solve this problem. This means that solving this problem optimally cannot be done in real-time. Therefore, we adopt the Genetic Algorithm (GA) to solve the problem in a computationally efficient manner, as detailed in the following section.
%
%
\section{Simulation Results}
\label{sec:Results}
In this section, we present and discuss the simulation results to evaluate the performance of the proposed CF-mMIMO-based joint power allocation and handover management (CF-JPAHM) technique, and we compare it with baseline techniques.

For the simulation, we consider a $1000\times 1000~\text{km}^2$ area that is covered by a cluster of $M$-LEO satellites. A set of UTs distributed uniformly are connected to the satellites as a group, for the CF-JPAHM technique, or based on single-satellite connection, for the baseline techniques. For the latter, we compare the performance of CF-JPAHM with that of a single satellite connection, where every UT is connected to the satellite with the best channel condition \cite{wu2016graph}. This baseline technique is referred to as \texttt{BestChannel}. The second baseline technique is where every UT minimizes its handover rate by staying connected to the same satellite, as long as its achievable rate is greater than the minimum acceptable one; and switches to another satellite only when the rate falls below the threshold, $R_k^{\text{min}}$, \cite{wu2016graph}.  We refer to this baseline scheme as \texttt{MaxServTime}. For CF-JPAHM, we use the GA to solve the multi-objective optimization problem in (\ref{eq:OP1}). The values of the simulation parameters are summarized in Table \ref{tab:simParameters}.

\begin{table}[t]
\renewcommand{\arraystretch}{1.3}
\caption{Simulation Parameters}
\label{tab:simParameters} 
\centering
\begin{tabular}{llc}
\hline 
\textbf{Parameter} & \textbf{Value} \\ 
\hline 
Satellite altitude  & $550$ km \\
Antenna factor ($\eta$) & 20 \cite{li2020hierarchical}\\
Carrier frequency & $30$ GHz \\
Shadowing std & $5$ dB \\
Noise figure & $7$ dB \\
Noise power spectral density & $-174$ dBm/Hz \\
Sat. max power ($P_m^{\text{max}}$) & $15$ dBW\\
Sat. antenna gain & $30$ dB\\
UT antenna gain & $5$ dB \\
Pilot power ($q_k$) & $5$ dBW \\
Coherence intervals: $\tau_c,~\tau_u^p$ & $300,~30$ samples\\
Number of runs & $10$ \\
Priority factor ($\alpha$) & $0.5$ \\
\hline
\end{tabular}
\end{table}
\begin{figure*}[t!]
\begin{minipage}{0.5\textwidth}
 \centering
 \includegraphics[width=0.95\linewidth]{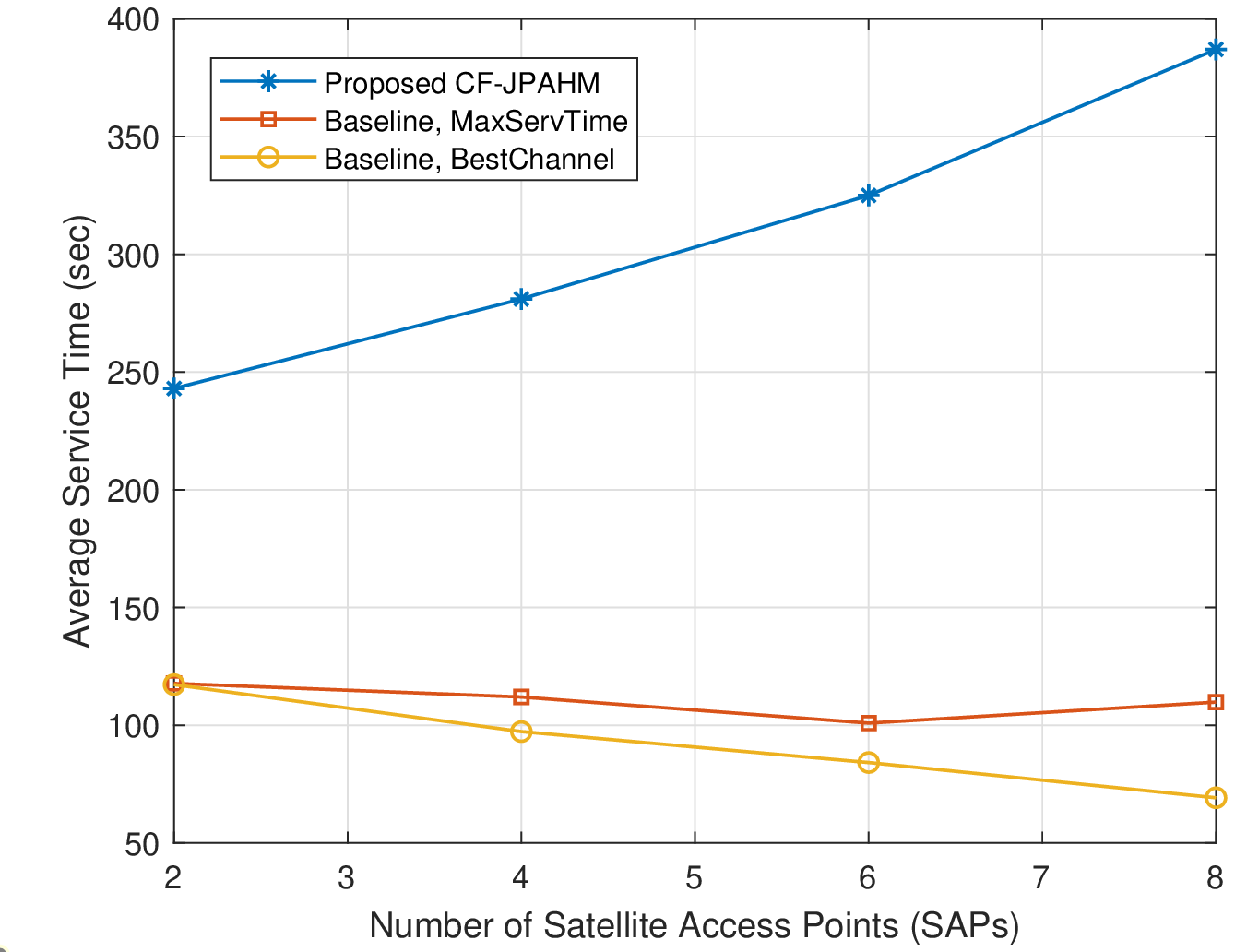}
 \caption{Average service time of the UTs.}\label{fig:AvrgeServTime}
\end{minipage}\hfill
\begin {minipage}{0.5\textwidth}
 \centering
 \includegraphics[width=0.95\linewidth]{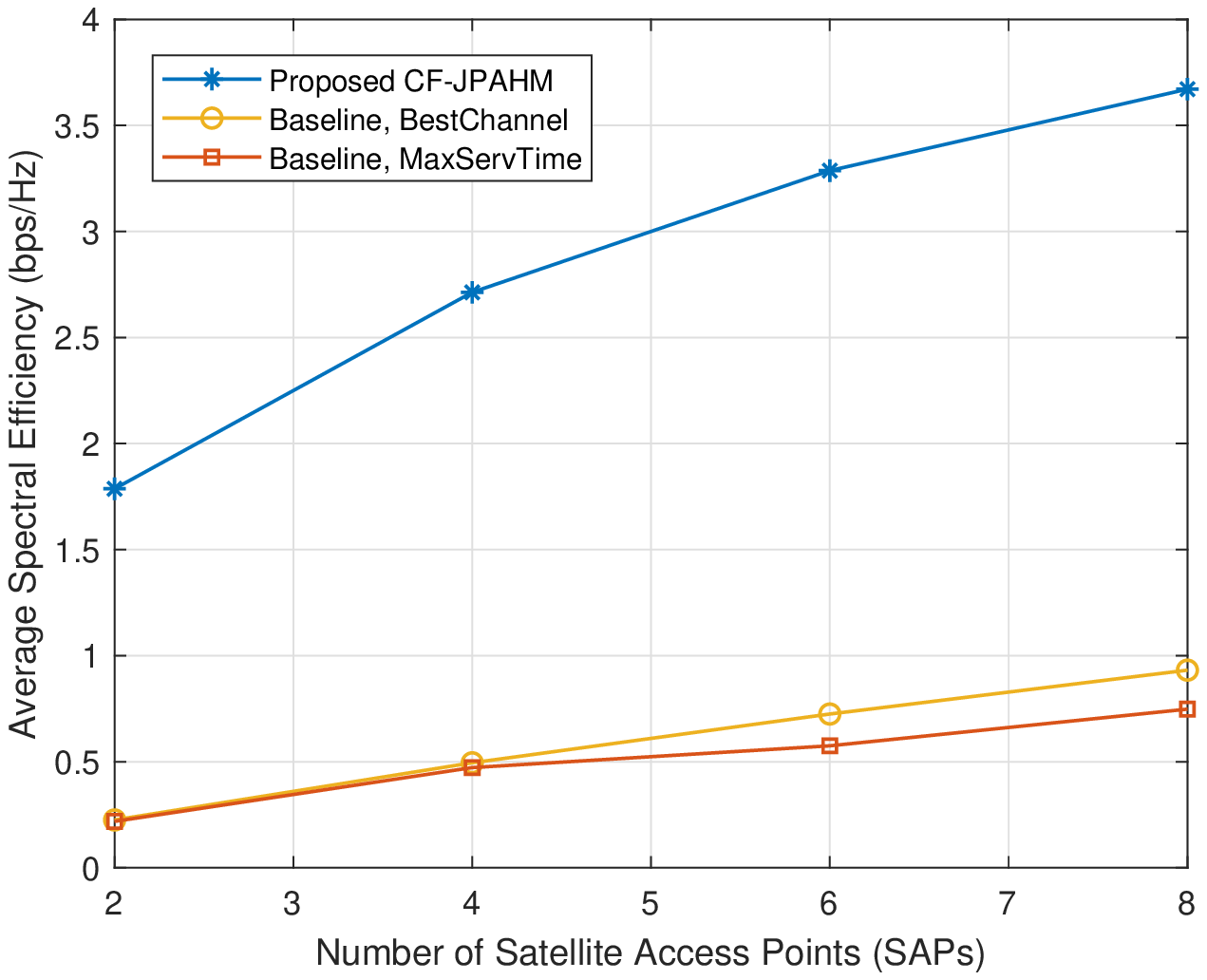}
 \caption{Average spectral efficiency of the UTs.}\label{fig:AvrgeSE}
\end{minipage}
\end{figure*}

Fig. \ref{fig:AvrgeServTime} shows the average service time of the UTs versus the number of SAPs based on the previously discussed techniques. As we can see, the proposed CF-JPAHM scheme achieves a higher average service time, and hence a lower handover rate for the UTs. This average service time increases by increasing the number of SAPs per cluster, for this allows more UTs to be served for a longer time. The baseline techniques have a lower average service time due to the connectivity to a single satellite at any instance of time. However, \texttt{MaxServTime} yields  better performance compared to the \texttt{BestChannel} technique since the latter switches between satellites for the best channel at any time, which reduces the average service time.

In Fig. \ref{fig:AvrgeSE}, the average spectral efficiency (in bps/Hz) is plotted against the number of SAPs using the different techniques and architectures considered. As we can see, the cooperative and optimized transmission of the satellites in the CF-JPAHM scheme can improve the spectral efficiency to a large extent. This improvement increases in relation to the number of SAPs. By contrast, the \texttt{BestChannel} and \texttt{MaxServTime} techniques are based on a single-satellite connectivity and achieve a lower spectral efficiency compared to the cell-free architecture.  
%
\section{Conclusions}
\label{sec:Conclusions}
In this paper, we introduced a novel CF-mMIMO-based architecture for future ultra-dense LEO SatNets. We investigated various network aspects, such as duplexing technique, pilot assignment, beamforming, and handover management. Then, we proposed a joint optimization framework for the power allocation and handover management processes to maximize the network throughput and minimize the handover rate while considering minimum QoS demands of the UTs. The simulation results demonstrated that the proposed architecture and optimization framework perform better than traditional techniques and can improve the performance of future LEO SatNets to a large extent.  
%
\ifCLASSOPTIONcompsoc
  \section*{Acknowledgments}
\else
  \section*{Acknowledgment}
\fi
This work has been supported by the National Research Council  Canada's (NRC) High Throughput Secure Networks program (CSTIP Grant  \#CH-HTSN-607) within the Optical Satellite Communications Canada (OSC)  framework.
%
%
%
\bibliographystyle{ieeetr}
\bibliography{References}
\end{document}